\documentstyle[aps,prl,epsf,color]{revtex}

\begin{document}

\unitlength1.0cm
\draft

\title{Transport in a laser irradiated thin foil}

\author{H. Ruhl}
\address{Max-Born-Institute, Max-Born-Stra\ss{}e 2a, 12489 Berlin, Germany} 

\date{\today}

\maketitle

\begin{abstract}
Three dimensional Particle-In-Cell simulations describing the
interaction of a short intense laser pulse with thin foils are
presented. It is observed that the laser generated electron current 
decays into magnetically isolated filaments. The filaments grow in 
scale and magnitude by magnetic reconnection. Two different laser 
wavelengths are considered. The spatial separation of the filaments 
varies for the two wavelengths. Many current filaments carry net 
electric currents exceeding the Alfven current considerably.
\end{abstract}
\pacs{52.40.Nk}


Key issues of laser-matter interaction at high intensities are the absorption 
of large fractions of the irradiated laser energy and the transport of 
large energy flows through a plasma. A typical application of intense 
laser-matter interaction is Fast Ignition (FI) in Inertial Confinement 
Fusion (ICF) \cite{TabakPHP94,HainPRL01,RothPRL01}. While FI requires 
large laser facilities high field experiments with thin foils promise 
to show interesting transport properties in dense plasma with moderate 
requirements for pulse energy \cite{RuhlPRL99,VshivkovPHP98}. Laser heated 
foils have been proposed for x-ray generation and harmonic emission 
\cite{DzhidzhoevJOSAB96,GizziPRL96}, for the production of well-defined 
dense plasma films \cite{ForsmanPRE98}, and as targets for efficient
ion acceleration \cite{RuhlPPR01}. Hence, there is both experimental and
theoretical interest in a good characterization of laser-irradiated 
thin foils.

We call foils thin when their thickness is larger than a few skin lengths
$l_{\text{s}}=c/\omega_{\text{p}}$, where $\omega_{\text{p}}$ is the plasma
frequency, and very much smaller than the mean free path of the electrons.
This definition has some ambiguity since a reasonable definition depends 
on the physics which is investigated. The latter however is not fully known. 
Two dimensional (2D) simulations in the plane perpendicular to the laser 
direction have recently been reported \cite{MTVPRL00}. In these simulations 
the laser has been neglected and fast particles have been injected by hand.
However, it is an open question to which extend laser-generated spectra
can be modeled. In addition, 2D geometry severely limits the available 
degrees of freedom for current transport and magnetic field evolution. 

Here we investigate mechanisms of self-organization in laser-generated 
charge flows in thin foils in full 3D where the laser produces the hot 
particle spectra. We show that merging of magnetic filaments 
\cite{Rosenbluth} is an important process to build up larger filaments. 
We show that this process proceeds slowly in time as soon as a
balance between magnetic and thermal pressures in the plasma is reached.
For a relatively cold plasma into which a well defined beam is injected
it is found that the Alfven current \cite{AlfvenPR39,LawsonJEC58} 
cannot be exceeded \cite{MTVPRL00,HondaPHP00}. However, in this paper it 
is revealed that laser generated charge flows can exceed a reasonably 
defined Alfven current limit. To address the problems discussed we perform 
two Particle-In-Cell simulations (PIC) with different parameters in 3D for 
a short intense laser pulse interacting with a sharp edged thin foil of 
many times over-critical plasma. Collisions will be neglected in both 
simulations.

The simulations mentioned differ only in the laser wavelengths. The first 
one has $\lambda=0.815 \mu \text{m}$ and the second $\lambda=1.612 \mu 
\text{m}$. All fields depend on $x$, $y$, and $z$. The simulation boxes 
are $10 \mu \text{m} \times 10 \mu \text{m} \times 4 \mu \text{m}$ large
and periodic in lateral directions. The grids have $400 \times 400 \times 
400$ cells. Electrons and ions are presented by equal quasi-particle numbers. 
The total number of quasi-particles in the simulations is $1.28 \cdot 10^8$. 
The initial electron and ion temperatures are $10.0 \, \text{keV}$ and 
$1.0 \, \text{keV}$ respectively. The laser pulses propagate in $z$ and 
are linearly polarized along $x$. After a rise time of three optical cycles 
the laser intensity is kept constant. The incident laser pulses have a 
Gaussian envelope laterally with a width of $2.5 \mu \text{m}$ at 
full-width-half-maximum. The irradiances in both cases are $I\lambda^2=6.0 
\cdot 10^{18} \text{Wcm}^{-2}\mu \text{m}^2$. The foils have a marginal 
initial deformation to enhance absorption to generate large currents
\cite{RuhlPRL99}. The deformations are parameterized by $z(x,y)=\delta 
\cdot \exp \left( -(x-x_{\text{0}})^2/r^2+(y-y_{\text{0}})^2/r^2 \right)$ 
where $\delta=0.4 \mu\text{m}$, $x_{\text{0}}=y_{\text{0}}=5.0 \mu
\text{m}$, and $r=2.5 \mu\text{m}$. The center of the foils are located 
at $z=2.2 \mu \text{m}$. The thicknesses are $0.6 \, \mu\text{m}$. The 
background ions consist of carbon and are assumed to be singly charged. 
The plasma density in the simulations is $n_{\text{e}}=3.33 \cdot 10^{22} 
\, \text{cm}^{-3}$. The coordinate system used in the simulations is right 
handed. Hence, the positive $z$-axis points out of the $xy$-plane shown in 
the figures. In what follows we will always use the positive $z$-axis for 
reference. 

Plot (a) of Fig. \ref{fig:bt} shows the plane $y=5.0 \, \mu \text{m}$ 
of the cycle averaged magnetic field $|{\bf B}|$ defined in the figure 
caption. Plot (b) of the same figure gives the plane $z=2.6 \, \mu 
\text{m}$ of $|{\bf B}|$. Plot (c) shows a magnification of the central 
area of the foil in plot (b) at $z=2.6 \, \mu \text{m}$. The arrows 
plotted over the pictures indicate the direction of the field vector 
of ${\bf B}$. Plots
(a,b,c) are for $\lambda=0.815 \mu\text{m}$ while plots (d,e,f) show the
same for $\lambda=1.62 \mu\text{m}$. In the center of the foil magnetic 
filaments are observed. As is seen from plots (c,f) the magnetic field 
in the filaments rotates counter-clockwise when viewed along the positive 
$z$-axis. The peak magnetic field strength which is obtained in the 
filaments is about $8700 \, \text{T}$ for $\lambda=0.815 \mu\text{m}$
and $5500 \, \text{T}$ for $\lambda=1.63 \mu\text{m}$. The white arrows 
in plots (b,e) show that the central magnetic filaments are surrounded 
by a magnetic fields of larger scale. The topology of this field is such 
that it rotates clockwise in front of the foil where $z < 2.2 \mu \text{m}$ 
holds and counter-clockwise at the rear of it for $z > 2.2 \, \mu \text{m}$. 
This is a consequence of electrons escaping into the vacuum both at the 
front and rear of the foil in opposite directions. The electrons escape 
from the center and return along the front and rear foil surfaces. The 
total electric current obtained by integrating over the whole lateral 
foil extension disappears on both the front and rear sides of the foil 
(see Fig. \ref{fig:currents}).

Plots (a,b,c) of Fig. \ref{fig:btevol} show the evolution of the 
magnetic filaments for $\lambda=1.63 \mu\text{m}$. Plot (a) indicates 
the early stage of the filament evolution. Filaments are very small 
initially. There have been efforts to describe this early stage with 
the help of the Weibel theory \cite{CalifanoPRE97,SentokuPHP00}.
Later the small filaments reconnect \cite{Rosenbluth} to form larger 
scales as is seen from plot (b). Plot (c) shows the saturation scale. 
Reconnection in the present context can be understood qualitatively 
with the help of Eq. (\ref{pressure_balance_integral}) derived later
in the paper which describes the total force on a finite plasma volume 
$V$. In case $V$ contains 
a single filament for which ${\bf j} \times {\bf B}$ has a large degree 
of asymmetry a net force is obtained. When two filaments approach 
each other the magnetic fields of the latter tend to cancel in the 
overlapping region leading to further asymmetry in ${\bf j} 
\times {\bf B}$. Hence, the two filaments reconnect to form a larger 
one. The magnetic field topology is changing in the course of the 
process since initially closed field lines in different filaments
end up as a single new filament with again closed field lines. The 
energy required is provided by the laser driver. 

Plot (a) of Fig. \ref{fig:jzt} shows the cycle averaged current filaments 
obtained in the foil for $\lambda=0.815 \mu\text{m}$. The simulation 
reveals roughly twenty of them where only the dark dots visible in the
center of plot (a) have been counted for a filament. Plot (b) of the same 
figure gives a magnified view of the central current filaments for better
illustration. The peak current density in the filaments is $|j_{\text{z}}|
=1.6 \cdot 10^{17} \, \text{A/m}^2$. Plots (c,d) show the same for $\lambda
=1.62 \mu\text{m}$. The peak current density in the filaments is now 
$|j_{\text{z}}|=7.6 \cdot 10^{16} \, \text{A/m}^2$. It is evident that the 
filament separation and scale are larger for $\lambda=1.62 \mu\text{m}$ 
while the current density is approximately half the short wavelength value.
Current density and filament scale are correlated as is implied by Eq.
(\ref{pressure_balance}) derived later in this paper.

Plots (a,b,c) of Fig. \ref{fig:alfven} compare current density,
total current, and Alfven current in the central region of the
foil for $\lambda=1.62 \mu\text{m}$. The white rectangles indicate 
the area over which the current density in (a) has been integrated 
to obtain the total current given in (b). The size of the rectangle
is such that it matches approximately the cross sectional area of
a single filament. Plot (b) has been obtained moving the rectangle 
over plot (a) point-wise. The Alfven current shown in (c) has 
been calculated with the help of $I_{\text{A}}=17.0 \, \beta \gamma 
\, \text{kA}$, where $\beta=v/c$ and $\gamma=1/\sqrt{1-\beta^2}$. 
The quantity $\beta$ has been obtained from the averaged velocity 
obtained from the fastest particles that constitute the net current
in a particular space-time point. This means that $c\beta({\bf x},t)=
\int_{fast} d^3p \, v_{\text{z}} \, f({\bf x},{\bf p},t)/ \int d^3p \, 
f({\bf x},{\bf p},t)$. Next, $\beta({\bf x},t)|_{\text{z}=2.6 \mu 
\text{m}}$ is averaged over the cross sectional area indicated
by the white rectangle shown in plot (c). It is seen that the Alfven 
current limit thus obtained is exceeded substantially by the net current
present in the highlighted current filament. We note that there are 
large ambiguities about how to calculate the Alfven current limit in 
a hot laser plasma since no unique particle beam can be identified in
the particle spectrum. We note further that in our simulations fast 
electrons do not contribute significantly to the net charge flow. The 
latter is predominantly due to slow electrons. As a consequence the 
averaged $\beta$ is small.

Plot (a) of Fig. \ref{fig:currents} shows the electric and magnetic
field energy distributions for both simulations obtained from the 
cycle averaged fields ${\bf E}$ and ${\bf B}$ by integrating the field 
energy densities laterally over the whole foil extension. In the vacuum 
the electric field energy dominates. In the foil the magnetic field 
energy dominates while the cycle averaged electric field disappears. 
To show that the total current and return current balance while locally 
they differ substantially plot (b) of the same figure shows the current 
in both foils obtained by integrating $j_{\text{z}}$ laterally as in 
plot (a). The peak values of the total electric currents are approximately 
$I=5.0 \cdot 10^5 \, \text{A}$ for $\lambda=0.815 \mu\text{m}$ and $I=2.5 
\cdot 10^5 \text{A}$ for $\lambda=1.62 \mu\text{m}$. For comparison,
the total currents limited by the Alfven current are $7.6 \cdot 10^4 
\text{A}$ for $\lambda=0.815 \mu\text{m}$ and $4.1 \cdot 10^4 \text{A}$ 
for $\lambda=1.63 \mu\text{m}$.

To derive a relation between the magnetic force acting on a filament 
and the thermal pressure inside the foil we neglect the electric field 
${\bf E}$ since it disappears there. In addition, we are only interested 
in slowly varying time scales. Under these assumptions we find making 
use of the Vlasov equation

\begin{eqnarray}
\label{pressure_balance}
\left( {\bf j} \times {\bf B} \right)_i&=&\partial_j P_{ij} \; , \qquad
P_{ij}=\int d^3p \; p_i v_j f \; .
\end{eqnarray}

Summation over repeated indices is implied. The quantity $P_{\text{ij}}$ 
denotes the pressure tensor which is obtained directly from the simulations. 
The integral form of Eq. (\ref{pressure_balance}) is given by

\begin{eqnarray}
\label{pressure_balance_integral}
\int_{V} d^3x \; \left( {\bf j} \times {\bf B} \right)_i
&=&\int_{\partial V} dS_j \, P_{ij} \; .
\end{eqnarray}

According to Eq. (\ref{pressure_balance_integral}) the force
on the plasma volume $V$ is obtained by the pressure acting on
the surface surrounding the volume. From the simulations we 
find that $|\partial_{\text{j}} P_{\text{zj}}| \ll |\partial_{\text{j}} 
P_{\text{xj}}|, |\partial_{\text{j}} P_{\text{yj}}|$ holds.
This is consistent with the observation that $|{\bf j} \times 
{\bf B}|_{\text{x,y}} \gg |{\bf j} \times {\bf B}|_{\text{z}}$ 
holds inside the foil. Figure \ref{fig:pressure} shows magnitude and 
direction of $\partial_{\text{j}} P_{\text{ij}}$ obtained from 
the simulations. Multiplying the magnetic field obtained from 
plot (f) of Fig. \ref{fig:bt} with the current density $j_{\text{z}}$ 
in plot (d) of Fig. \ref{fig:jzt} and comparing with Fig. 
\ref{fig:pressure} it is seen that the magnetic force on the current 
is approximately balanced by the kinetic pressure of the electrons 
for the quasi-steady state shown in the present paper.

Fig. \ref{fig:spectrum} shows the electron spectrum obtained for 
$\lambda=1.62 \mu \text{m}$. We find that current and return current 
contain fast electron populations. The fast electrons in the return 
current are fast electrons that have been reflected at the back surface 
due to the electric field present there. Comparing plots (a,b) shows 
that the fast electrons in the return current try to escape from the 
filaments. Peak electron energies obtained for $\lambda=1.63 \mu 
\text{m}$ are of the order of $1.0 \, \text{MeV}$ as is seen from 
plots (a,b) while the electron temperature of the hot electrons found
within the inner rectangle (see plot (c) of Fig. \ref{fig:spectrum}
and plot (a) of Fig. \ref{fig:alfven}) as well as within the inner and 
outer rectangles (see plot (d) of Fig. \ref{fig:spectrum} and plot (a)
of Fig. \ref{fig:alfven}) is roughly $300.0 \, \text{keV}$. To understand 
these low values we note that the foil represents a sharp edged plasma 
with $n/n_{\text{c}}=80.0$ for $\lambda=1.63 \mu \text{m}$. Hence, the 
laser radiation decays rapidly over distances of $l_{\text{s}} \approx 
0.03 \mu \text{m}$. As a consequence the effective fields in the plasma 
capable of accelerating electrons become very small. 

Transport properties in thick foils cannot be directly predicted on 
the basis of those found in thin foils. The main reason is that the 
electron distribution function in thin foils tends to become symmetric 
in momentum space which is not so for extended plasma slabs. As is 
found in this paper, aa approximate force balance between thermal 
pressure and magnetic force can be established in thin foils which 
is capable of slowing down further filament reconnection. It needs 
to be investigated two which degree the latter is true in thick plasma 
slabs.

In conclusion, we have shown with the help of 3D PIC simulations
that large charge flows generated by intense laser radiation in 
thin foils decay into current filaments. The latter originate from the 
critical surface and extend into the bulk plasma. Their separation 
and magnitude is correlated with the laser wavelength. The current 
filaments are surrounded by magnetic filaments. The evolution of 
the magnetic filaments is mainly governed by filament merging. In 
our simulations this process slows down when an approximate balance 
between thermal pressure and the magnetic force acting on the 
filaments is reached. The net charge flow in a single filament can 
exceed a reasonable definition of the Alfven current substantially. 
For the short wavelength case studied in this paper the net current 
in a single filament exceeds the Alfven current up to $8$-fold. 
However, a reasonable Alfven current definition turns out to be 
ambiguous in the context of hot laser plasmas. 

This work has been sup\-ported by the Euro\-pean Com\-mission 
through the TMR network SILASI, contract No. ERBFMRX-CT96-0043.  
Use of the Supercomputing facilities at ZIB (Konrad Zuse, Berlin,
Germany) and NIC (John von Neumann Institute, J\"ulich, Germany) 
has been made.

\newpage

\begin{figure}
\caption[]{Cycle averaged ${\bf B}$. Plot (a) shows the plane $y=5.0 \, 
\mu \text{m}$, plots (b,c) the planes $z=2.6 \, \mu \text{m}$ of the 
cycle averaged magnetic field $B=(B^2_{x}+B^2_{y}+B^2_{z})^{0.5}$ for 
$\lambda=0.81 \mu\text{m}$ and $t=54 \, \text{fs}$. Plots (d,e,f) show 
the same for $\lambda=1.62 \mu \text{m}$ and $t=130 \, \text{fs}$. The 
arrows indicate the direction of the cycle averaged magnetic field 
${\bf B}$. The parameter is $B_0=8.76 \cdot 10^{2} \, \text{Vs/m}^{2}$.}
\label{fig:bt}
\end{figure}

\begin{figure}
\caption[]{filament merging in ${\bf B}$. The plots show the planes 
$z=2.6 \mu\text{m}$ of the cycle averaged magnetic field $B=(B^2_{x}
+B^2_{y}+B^2_{z})^{0.5}$ for $\lambda=1.62 \mu\text{m}$ and $t=30 
\text{fs}$ (a), $t=40 \text{fs}$ (b), and $t=108 \, \text{fs}$ (c). The 
parameter is $B_0 = 8.76 \cdot 10^{2} \, \text{Vs/m}^{2}$.} 
\label{fig:btevol}
\end{figure}

\begin{figure}
\caption[]{Current density $j_{\text{z}}$. Plots (a,b) show the planes 
$z=2.6 \, \mu \text{m}$ of the cycle averaged electron current density 
$j_{\text{z}}$ for $\lambda=0.81 \mu\text{m}$ and $t=54 \, \text{fs}$. 
Plots (c,d) show the same for $\lambda=1.62 \mu\text{m}$ and $t=130 \, 
\text{fs}$. The arrows indicate the direction of the cycle averaged 
magnetic field ${\bf B}$. The parameter is $j_0=1.66 \cdot 10^{16} 
\text{A/m}^{2}$.}
\label{fig:jzt}
\end{figure}

\begin{figure}
\caption[]{Alfven current $I_{\text{A}}$. The plots show the planes 
$z=2.6 \mu\text{m}$ of the cycle averaged current density $j_{\text{z}}$ 
(a), the total current $I$ obtained from integrating $j_{\text{z}}$ over 
the white square (b), and the Alfven current $I_{\text{A}}$. The actual
current exceeds the Alfven current $4$-fold. The parameters are $j_0=1.66 
\cdot 10^{16} \text{A/m}^{2}$, $\lambda=1.62 \mu\text{m}$, and $t=130 
\text{fs}$.} 
\label{fig:alfven}
\end{figure}

\begin{figure}
\caption[]{Field energy and total current $I$. Plot (a) shows the 
electromagnetic field energies per micron in the foils calculated 
from the time averaged fields. Dotted lines correspond to $\lambda=0.81 
\mu\text{m}$ and $t=54 \text{fs}$. Solid ones belong to $\lambda=1.62 
\mu\text{m}$ and $t=130 \text{fs}$. Bold lines give the magnetic field 
energy for both cases which has been multiplied by a factor of five. 
Plot (b) shows the electron current $I_{\text{z}}$. Dotted lines 
correspond to $\lambda=0.81 \mu\text{m}$ and $t=54 \text{fs}$. Solid 
ones belong to $\lambda=1.62 \mu\text{m}$ and $t=130 \text{fs}$.}
\label{fig:currents}
\end{figure}

\begin{figure}
\caption[]{Pressure gradient. Cycle averaged thermal force density 
$|\partial_{\text{j}} P_{\text{i,j}}|$. The colorbar is in units 
$j_{\text{0}}B_{\text{0}}=1.46 \cdot 10^{19} \text{N/m}^3$ for
comparison with Figs. \ref{fig:bt} and \ref{fig:jzt}. The white 
arrows show the force direction. The parameters are $\lambda=1.62 
\mu \text{m}$ and $t=130 \text{fs}$.}
\label{fig:pressure}
\end{figure}

\begin{figure}
\caption[]{Electron spectrum. Plot (a) shows the $zp_{\text{z}}$-plane 
of the phase space located inside the inner rectangle indicated in plot (a)
of Fig. \ref{fig:alfven}. Plot (b) shows the same for particles
located between the inner and outer rectangles in plot (a) of Fig. 
\ref{fig:alfven}. Plot (c) shows the energy spectrum obtained
form the particles located in the inner rectangle while plot (d) 
shows the one obtained from electrons between the inner and outer
rectangle. The parameters are $\lambda=1.62 \mu \text{m}$ and $t=130 
\text{fs}$.}
\label{fig:spectrum}
\end{figure}









\begin{references}

\bibitem{TabakPHP94} M. Tabak {\em et al.}, Phys. Plasmas {\bf 1}, 1626
(1994).

\bibitem{HainPRL01} S. Hain and P. Mulser, Phys. Rev. Lett. {\bf 86},
1015 (2001).

\bibitem{RothPRL01} M. Roth {\em et al.}, Phys. Rev. Lett. {\bf 86},
437 (2001).

\bibitem{RuhlPRL99} H. Ruhl {\em et al.}, Phys. Rev. Lett. {\bf 82},
2095 (1999).

\bibitem{VshivkovPHP98} V. Vshivkov {\em et al.},  Phys. Plasmas 
{\bf 5}, 2727 (1998).

\bibitem{DzhidzhoevJOSAB96} M. S. Dzhidzhoev {\em et al.}, 
J. Opt. Soc. Am. {\bf B 13}, 143 (1996).

\bibitem{GizziPRL96} L. A. Gizzi {\em et al.}, Phys. Rev. Lett.
{\bf 76}, 2278 (1996).

\bibitem{ForsmanPRE98} A. Forsman {\em et al.}, Phys. Rev. E {\bf 58},
R1248 (1998). 

\bibitem{RuhlPPR01} H. Ruhl {\em et al.}, accepted for publication
in Plasma Physics Reports, May (2001).

\bibitem{MTVPRL00} M. Honda {\em et al.}, Phys. Rev. Lett. {\bf 85},
2128 (2000).

\bibitem{Rosenbluth} A. A. Galeev, Handbook of Plasma Physics, ed. 
M. N. Rosenbluth and R. Z. Sagdeev, ISBN: 0444866450, North Holland 
Amsterdam, 305 (1984).

\bibitem{AlfvenPR39} H. Alfven, Phys. Rev. {\bf 55}, 425 (1939).

\bibitem{LawsonJEC58} J. D. Lawson, J. Electron Control {\bf 5}, 
146 (1958).

\bibitem{HondaPHP00} M. Honda, Phys. Plasmas {\bf 7}, 1606 (2000).

\bibitem{CalifanoPRE97} F. Califano {\em et al.}, Phys. Rev. E {\bf 56},
963 (1997).

\bibitem{SentokuPHP00} Y. Sentoku {\em et al.}, Phys. Plasmas {\bf 7},
689 (2000).

\end{references}
\end{document}